\documentclass[twoside]{article}

\usepackage[sc]{mathpazo} 
\usepackage[T1]{fontenc} 
\usepackage{microtype} 

\usepackage[english]{babel} 

\usepackage[hmarginratio=1:1,top=32mm,columnsep=20pt]{geometry} 
\usepackage[hang, small,labelfont=bf,up,textfont=it,up]{caption} 
\usepackage{booktabs} 

\usepackage{lettrine} 

\usepackage{enumitem} 
\setlist[itemize]{noitemsep} 

\usepackage{color}

\usepackage{abstract} 

\usepackage{titlesec} 
\titleformat{\subsection}[block]{\large\itshape}{\thesubsection}{1em}{} 

\usepackage{fancyhdr} 
\usepackage{titling} 

\usepackage[hyphens]{url}

\usepackage[anythingbreaks]{breakurl}
\usepackage[breaklinks, colorlinks=true, citecolor=blue, filecolor=blue, linkcolor=blue, urlcolor=blue, pdfstartview=FitH]{hyperref} 

\usepackage{wrapfig}

\usepackage{natbib}

\usepackage{graphicx}

\usepackage{setspace}

\usepackage[utf8]{inputenc}
\usepackage[T1]{fontenc}

\addto\captionsenglish{}

\usepackage[final]{changes}

\pretitle{\begin{center}\Large\bfseries} 
\posttitle{\end{center}} 

\title{Usability Evaluation for Online Professional Search in the Dutch Archaeology Domain} 

\author{%
    \textsc{Alex Brandsen}\thanks{Corresponding author}\\[1ex]  
    \normalsize Leiden University \\ 
    \normalsize \href{mailto:a.brandsen@arch.leidenuniv.nl}{a.brandsen@arch.leidenuniv.nl} 
    \and 
    \textsc{Suzan Verberne}\\[1ex] 
    \normalsize Leiden University \\ 
    \normalsize \href{mailto:s.verberne@liacs.leidenuniv.nl}{s.verberne@liacs.leidenuniv.nl} 
    \and 
    \textsc{Karsten Lambers}\\[1ex] 
    \normalsize Leiden University \\ 
    \normalsize \href{mailto:k.lambers@arch.leidenuniv.nl}{k.lambers@arch.leidenuniv.nl} 
    \and 
    \textsc{Milco Wansleeben}\\[1ex] 
    \normalsize Leiden University \\ 
    \normalsize \href{mailto:m.wansleeben@arch.leidenuniv.nl}{m.wansleeben@arch.leidenuniv.nl} 
}
\date{} 
\renewcommand{\maketitlehookd}{%
\begin{abstract}

	\noindent This paper presents AGNES, the first information retrieval system for archaeological grey literature, allowing full-text search of these long archaeological documents. 
	This search system has a web interface that allows archaeology professionals and scholars to search through a collection of over 60,000 Dutch excavation reports, totalling 361 million words. We conducted a user study for the evaluation of AGNES's search interface, with a small but diverse user group. The evaluation was done by screen capturing and a think aloud protocol, combined with a user interface feedback questionnaire. The evaluation covered both controlled use (completion of a pre-defined task) as well as free use (completion of a freely chosen task). The free use allows us to study the information needs of archaeologists, as well as their interactions with the search system. We conclude that: (1) the information needs of archaeologists are typically recall-oriented, often requiring a list of items as answer; (2) the users prefer the use of free-text queries over metadata filters, confirming the value of a free-text search system; (3) the compilation of a diverse user group contributed to the collection of diverse issues as feedback for improving the system. We are currently refining AGNES's user interface and improving its precision for archaeological entities, so that AGNES will help archaeologists to answer their research questions more effectively and efficiently, leading to a more coherent narrative of the past.
 
    \noindent \textbf{Key Words:} Information Retrieval, Usability Evaluation, User Interface Evaluation, Text Mining
  
\end{abstract}
}

\begin{document}

\pagenumbering{gobble}

\maketitle
\section{Introduction}
\label{sec:intro}

	Archaeologists create large amounts of texts. Besides scholarly publications, another large source of documents are unpublished technical fieldwork reports. These reports are required to be produced by law whenever an excavation is performed \citep{CouncilofEurope1992EuropeanRevised}. They are generally not published in the traditional sense, and end up in various repositories, either in hard copy or digital format. The information in these reports is often needed, and described as `crucial' and `essential' by European archaeologists in a user study in the ARIADNE project \citep{Selhofer2014D2.1:Needs}. \deleted{In the Netherlands, a} A recent report by \cite{Habermehl2019OverOnderzoek} \replaced{states}{mentions} that the accessibility, findability, and searchability of research output is essential for synthesising research.
	
    In the Netherlands, the amount of reports created in the last twenty years is currently estimated at around 60,000, and is growing by approximately 4000 per year \citep{RCE2017DeErfgoedmonitor}. Most of these reports are categorised as `grey literature' \citep{Evans2015AParadoxes}, and are likely to end up in a proverbial `graveyard', unread and unknown, unless they are properly archived, indexed, and disclosed. 
	
	Easy access to this information is a major problem for the archaeological field, as there is currently no \replaced{free-text search system}{efficient search portal} available \added{for archaeological reports}. Searching through these documents and analysing them is a time consuming task when done by hand, and will often lack consistency \citep{Brandsen2019}. \deleted{Text mining and Information Retrieval (IR) provide methods for disclosing information in large text collections, allowing} \added{A full-text index of archaeological documents, with a user interface, would allow} researchers to locate (parts of) texts relevant to their research questions.

    Some studies have investigated applications of \replaced{Natural Language Processing (NLP)}{text mining} in heritage collections in general \citep{vanHooland2015ExploringCollections}, but also from archaeological reports specifically, both in English \citep{Vlachidis2017,Amrani2008AArchaeology,Byrne2010AutomaticText} and Dutch \citep{Paijmans2010,Vlachidis2017}. However no \added{IR} system is currently available that allows full-text access to the documents held in Dutch archives \citep{Habermehl2019OverOnderzoek}. As a result, relevant and valuable information is not being utilised at the moment.

    In this paper we present the AGNES search system that allows users to harness \added{IR and NLP} techniques to search for relevant archaeological literature\deleted{ (more information on this system in Section \ref{AGNES})}. To ensure that the needs of the potential users and stakeholders are met, \replaced{a focus group of archaeologists has been involved in the development and evaluation of the system}{this project includes a user study}. It is important that the usability of a system such as this is evaluated properly, as previous research indicates that there is a strong relationship between the usability and uptake of search systems \citep{Dudek2007IsEngines}.
    
    Archaeology is an archive-heavy discipline \added{in the digital humanities.} Much of \replaced{the archaeological}{our} data and finds reside in repositories. Yet to the best of our knowledge, no detailed research has been done into the information needs of archaeologists, nor of the usability of online tools for archaeology. 

    The following research questions are addressed in this paper: 
    
    \begin{enumerate}
        \item What type of information needs do archaeologists have?
        \item What are their query strategies?
        \item How satisfied are the users with the usability of the AGNES system?
        
    \end{enumerate} 
    
    The contributions of this paper in comparison with previous work
    is that this is (to our knowledge) the first full text search system and the first usability evaluation of such as system in the archaeology domain. We also investigate archaeologists' information needs, \added{their query strategies}, and evaluate the usability of our search system \added{for answering their information needs}. 
    
    The structure of the rest of this paper is as follows: Section~\ref{sec:background} provides an overview of related and prior work; Section~\ref{AGNES} is a short introduction to the current version of our system; Section~\ref{userstudy} presents the set up of the user study with the results presented in Section~\ref{sec:results}, followed by a discussion in Section~\ref{sec:discussion}. Section~\ref{sec:conclusions} describes our conclusions and future work.

\section{Background}
\label{sec:background}

    \subsection{Access to archaeological data}
    
        In Dutch archaeology, a number of \replaced{professional search systems}{digital archives} are currently used to \replaced{access}{find} excavation reports. The main two are EASY \citep{DANS2019DANSEASY} maintained by DANS (Data Archiving and Networked Servies) and Archis \citep{RijksdienstvoorhetCultureelErfgoed2019Archis} by the State Service for Heritage (RCE). The Dutch National Library (KB) also makes a limited amount of reports available via a standard library portal, but this system is used to a much lesser extent, due to the small amount of texts and the search interface not being geared towards archaeology. None of these systems support full text search, a highly desirable feature we \replaced{have included}{are including} in AGNES.
        
        This kind of search through archaeological reports is a form of professional search, which implies that the developed search interface is used by a specific group of professionals, as opposed to web search engines designed for the general public (e.g. Google). Professional search often has very specific user needs that go beyond the needs of the general public.

        In the ARIADNE project \citep{Niccolucci2013ARIADNE:Europe}, interviews and an online questionnaire were used to assess the current state of archaeological data access across Europe. Regarding problems encountered while searching for data, `most comments related to the accessibility of data. Data appeared as difficult to find, not available online, and if online difficult to access' \citep[][p. 63]{Selhofer2014D2.1:Needs}. Also, 93\% of respondents indicated that a portal enabling innovative and more powerful search mechanisms would be `very helpful' or 'rather helpful' \citep[][p. 63]{Selhofer2014D2.1:Needs}.
        
        More specifically for the Netherlands, \cite{Hessing2013EvaluatieEvaluatie} did an evaluation of the (then) current archaeological search systems in 2013. They found that the Archis system did allow for geographical search, but due to free text fields in the metadata forms, it is difficult to find \added{the relevant} items and make sure the results are exhaustive. A more recent report by \citet{Habermehl2019OverOnderzoek} shows this is still the case: they state that the current search systems are not useful enough.
        
        Since the research by \citeauthor{Hessing2013EvaluatieEvaluatie} a new version of Archis has been released (3.0) which allows search across all metadata fields and the plotting of results on a map; something very important to archaeologists as all \replaced{their}{our} research has a strong geographical component. It also allows searching in a specific area plotted on a map, but this cannot be combined with text search in the metadata, only faceted search.
        
        The EASY system also offers text search, but again only on metadata. At the time of \citeauthor{Hessing2013EvaluatieEvaluatie}'s report, there was no mapping functionality, but due to this study this has since been added, and results can now be displayed on map. None of the systems offer full text search of the documents themselves, only of (combinations of) metadata. While metadata can be more specific and precise than full text (depending on who created the metadata), it is often incomplete and prone to errors, which makes a full text search \deleted{with text mining} highly desirable.
    
    \subsection{Feedback on existing systems from our user group} 

        Research done early in the AGNES project confirms the findings above. In the initial user requirement solicitation workshop, we asked our user group about their current search behaviour. This showed that most researchers use the DANS search functionality and find it not sufficient for their search needs, with most people having to manually search through individual documents to find information. The Archis system is used to a lesser degree, again mainly because the search functionality is not sufficient and is experienced as being difficult to use. Specifically, not being able to search through all the text, and no proper integration of a map (including searching specific areas) were noted as currently missing. Multiple participants explained that they create their own literature lists with keywords to be able to find materials previously accessed \citep{Brandsen2019}. 
        
        We also performed user requirement solicitation, and the user group had a clear need for geographic search, plotting results on a map, and faceted search 
        \citep{Brandsen2019}. These kinds of features are rarely needed in \replaced{open-domain web}{public} search. Specifically the combination of these three features with full text search is highly desired, but not currently offered by any of the search portals we are aware of in the Netherlands and abroad. 
        
    \subsection{Related work in usability studies}

        Usability studies assess the extent to which a system is easy and efficient to use, and how well users can reach their goals. In other words, usability is the overall usefulness of a product \cite{Rosenzweig2015}. 
    
        A common evaluation method in usability studies is to have users from the target audience use the software, and ask them to give feedback on the system. In usability studies for IR systems, the most used evaluation protocol is to provide the users with a number of information problems and ask them to solve these problems using the search system at hand. A questionnaire is used after the process to assess their satisfaction \citep{Spink2002AStudy,Behnert2017AAssessments,Rico2019EvaluatingApproach}.
        
        \label{thinkaloud}    
        Besides asking for feedback after the session, another commonly used method for getting feedback during the use of the software is the Think Aloud Protocol, as originally proposed by \citet{Lewis1982UsingDesign}, and more recently applied by e.g. \citet{Gerjets2011MeasuringData} and \citet{Hinostroza2018BrowsingProtocol}. Research by \citet{vanWaes2000ThinkingHypertext} shows that the combination of thinking aloud and recording the user behaviour is a useful observation method to collect data about the searching process, both on usability and cognitive aspects, which is confirmed by e.g. \citet{Verberne2016InformationResearch.} and \citet{Kirkpatrick2018UsingOnline}. 
        
        In digital humanities studies, usability evaluation of tools and services is seen as a key part of the research  \citep{Bulatovic2016UsabilityProcessb}, and is published and discussed in detail \citep[e.g.][]{Steiner2014EvaluatingApproach,Bartalesi2016UsabilityDanteSources,Hu2018UsabilityLibrary}. In archaeology specifically, usability studies are less routinely performed (or at least not often published), and seem to be limited to the fields of virtual reality and digital museums \citep{Karoulis2006UsabilityInterface,Pescarin2014EvaluatingStudy}\replaced{. One}{, although a} recent study by \citet{Huurdeman2020MoreContent} investigates search interface features for 3D content in a digital heritage context. 
        
        Giving that there are key limitations to the currently available \added{archaeological IR} systems, and usability studies are rare in the archaeology domain, we think it is vital to research and publish the usability of the system we are creating.

\section{AGNES}
    \label{AGNES}

    In the current project, we are developing AGNES, an IR system that makes Dutch archaeological grey literature more accessible and searchable\replaced{. The AGNES index currently contains }{ by applying Information Retrieval techniques to this large data set (}roughly 60,000 documents, totalling 361 million words\replaced{. The PDF documents are stored in}{, from} the DANS repository. 

    AGNES consists of three parts: software for recognising archaeological concepts (named entities), an indexing system that stores these entities and the full text, and a web application front end that can search through this index. 

    Named entities are terms that refer to important concepts from the real world  \citep{Marrero2013NamedOpportunities}. In the context of this project, the entities are archaeological concepts, mentioned in excavation reports. To give an example, in the following sentence the entities are bold: `The \textbf{burial mound} yielded some \textbf{scrapers} from the \textbf{Neolithic}', a context, an artefact, and a time period, respectively. The example illustrates that entities can consist of multiple words. Two particular challenges of entity recognition are that a single term can refer to multiple entity types (e.g. `Swifterbant' can be either a location, a time period, or a type of pottery), and that multiple terms can refer to the same entity (e.g. `Neolithic' and `New Stone Age'). For more technical information on the NER process, as well as the methods used, see \citet{Brandsen2020CreatingDomain}. 
    
    In AGNES, archaeological entities are recognised and labelled during the indexing of the documents. In version 0.3 of AGNES, all 60,000 reports from the DANS repository were indexed. For each page in these documents, the named entities are extracted and combined with the full text of the page and indexed directly by ElasticSearch \citep{Gormley2015Elasticsearch:Engine}\deleted{, an open source search engine running on a web server}. We are currently indexing at the page and document level, but in future we will index at the chapter/section level. This is more suitable to most information needs, as researchers will want to find e.g. all sections that mention `axe' and `neolithic', even if they are mentioned on different pages. This was also seen in the user study, as detailed in the next section.
    
    \begin{figure}[t]
        \centering
        \includegraphics[width=0.75\textwidth]{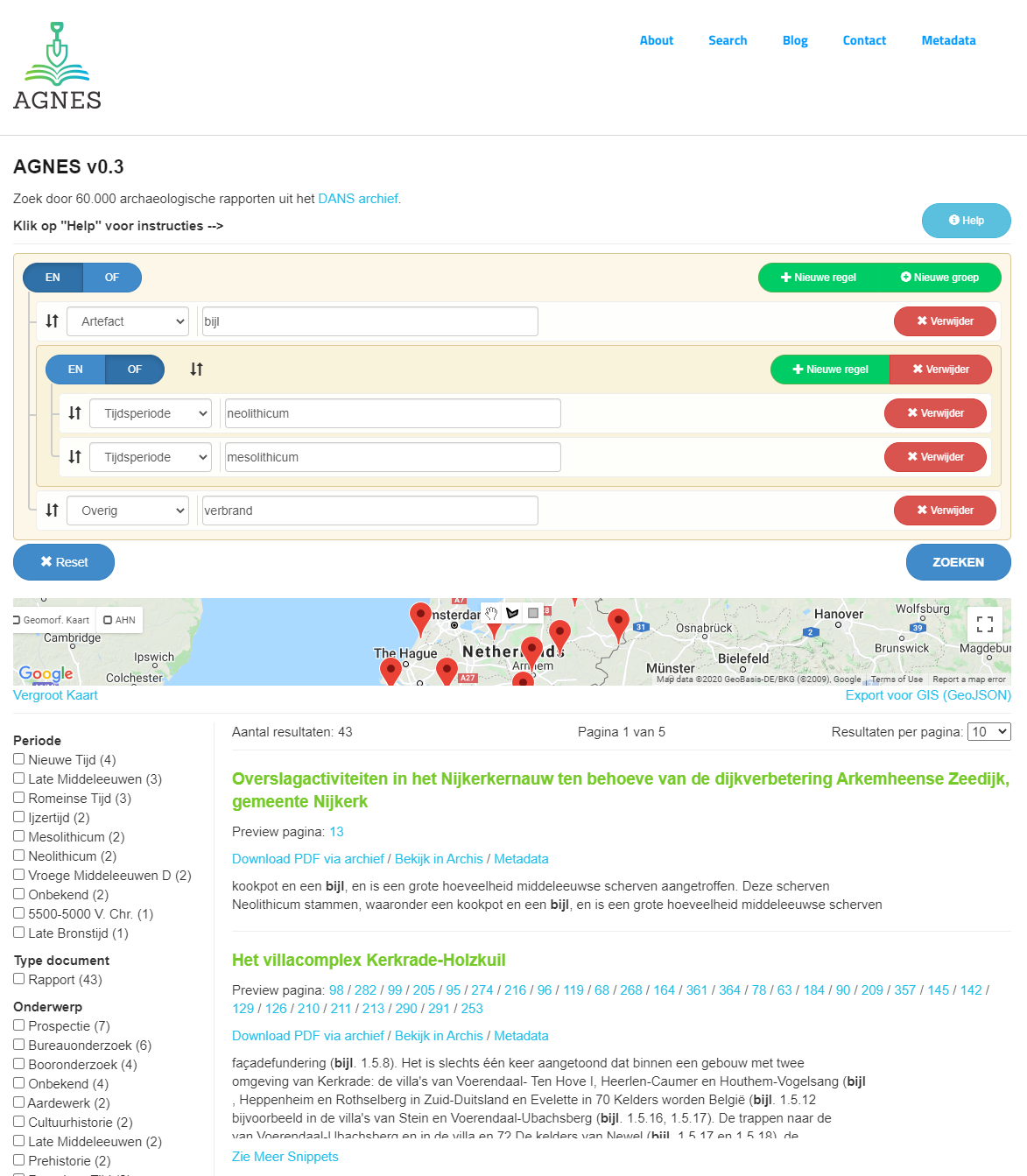}
        \caption{Screenshot of AGNES version 0.3. Pictured here is a query for `artefact:axe AND (period:neolithic OR period:mesolithic) AND fulltext:burnt', with the results on a map and in a list underneath (with snippets). On the left we can see the facets, used to filter results on period, type of document, and subject.}
        \label{v03}
    \end{figure}
        
    We developed a front end to query the index. The searcher can use a query builder \citep{Sorel2018JQueryQueryBuilder} that allows for boolean AND / OR logic. They can specify exactly which entity they are looking for in each part of the query, or select a general full-text search. This visual interface allows for \deleted{complex} the creation of queries such as the following pseudo-query\footnote{Note that this is not what the user types in, but an easy to read representation of the query that's generated by the system}: 
        
\begin{verbatim}
artefact:axe AND (period:neolithic OR period:mesolithic) AND fulltext:burnt
\end{verbatim}
        
    \noindent which returns results on axes from the Neolithic or Mesolithic where the word `burnt' is also mentioned on the same page. It is also possible to refine the query by using facets (filtering for specific metadata values, such as time period or document type) or by drawing a polygon on a map, performing a geographical search.

    The query is then sent to ElasticSearch, which returns a list of matching results. Once the results are displayed, the user can view a snippet of the text surrounding the keywords, preview the page of the report or go directly to the DANS repository to download the document. No PDFs are made available on the AGNES server in order to respect the copyright of these files.
    
    \deleted{In the second year of the project, AGNES version 0.3 has been completed, which implements changes to the UI as suggested by the user group in the initial workshop. }See Fig. \ref{v03} for a screenshot of the AGNES UI. This version is the one that has been evaluated in the current study, and is available at \url{http://agnessearch.nl/v03}.\footnote{Please note, free registration is needed to access the system.}

\section{User Study Setup}
    \label{userstudy}

    \label{focusgroup}

    A focus group is a small but diverse group of people who's reactions are studied to generalise to a larger population. Focus groups are often used for data collection, and have been studied and described in detail in literature \citep{Thomsett-Scott2006WebGroups,barbour2018doing}. Specifically, they are useful for gathering qualitative data quickly and cheaply, as well as gathering data on attitudes, values, and opinions  \citep{Cohen2002ResearchEducation}. This very much aligns with the purpose of this study; to collect users' opinions on the currently available systems, their requirements for a new system and their assessments of developed features.
 
    \subsection{Workshops in the AGNES project}
    
        For the user study, we followed a user-centred approach, consisting of pre-assessment (user requirement solicitation), mid-term evaluation (feedback on early system versions), and post-assessment (user trial). A user-centred evaluation approach focuses on examining the behaviour and preferences of users, and their interaction with the system. The main purposes of this type of evaluation are to find problems and assessing the quality of a system \citep{Dejong1997Reader-FocusedMethods}, exactly what we set out to do.
       
        Four workshops are held during the AGNES project, once per year. The first workshop had the aim of eliciting the requirements of the users, and the second workshop aimed to evaluate the user interface. Later workshops will focus more on assessing the quality of the results. Minutes are taken at each session to record the comments and feedback of the group, and these will be made public after anonymisation. 
 
    \subsection{Compilation of the focus group}
    
        \begin{table}[t] 
          \centering
          \begin{tabular}{|l|l|l|}
            \hline
            \textbf{Category}      & \textbf{Situation}  & \textbf{Count} \\ \hline
            Academia               & PhD Student         & 4              \\ \hline
            Academia               & Assistant Professor & 1              \\ \hline
            Academia               & Lecturer			 & 1              \\ \hline
            Commercial Archaeology & Excavation          & 1              \\ \hline
            Commercial Archaeology & Prospection         & 1              \\ \hline
            Government             & Municipal           & 1              \\ \hline
            Government             & National            & 1              \\ \hline
          \end{tabular}
          \caption{Overview of participants in usability evaluation per category}
          \label{participants-table}
        \end{table} 
        
        We compiled a focus group of archaeologists. The size and compilation of this group is fluid, and can be changed during the project to fit with the current goals and/or address issues of representativeness.
        
        This group has been selected in such a way that it includes every category of the target audience as defined in \citet{Brandsen2019}.
        At the current stage of research, the group consists of six academics, two commercial professionals, and two archaeologists working on different levels in government. See Table \ref{participants-table} for a more detailed break down of the participants. 
        
        Regarding the size of the focus group, \citet{Nielsen1993AProblems} show that the number of additional usability problems encountered when adding more users quickly decreases beyond five users. \replaced{Thus, }{So we believe} the current size of the focus group should be more  adequate in this regard.

    \subsection{Design and procedure}

        The evaluations were performed on a one-to-one basis\deleted{, in contrast to the initial group workshop}. Users only got an introduction to what the system was, but no specific instructions on how to use the system. We placed the users in a quiet office with the system running on a laptop, and asked each participant to use the system to perform three predefined tasks, as well as at least three of their own \replaced{self-defined information needs}{research questions}. The predefined tasks are the following (translated from Dutch):
        
        \begin{enumerate}
            \item Where in the Netherlands can we find globular jars (\textit{kogelpotten}) in fire pits?
            \item Find all literature relating to Neolithic scrapers found south of the river Meuse.
            \item Find all Roman pottery found in a settlement.
        \end{enumerate}
        
        The first task \deleted{is fairly simple, and} is intended to introduce the user to the query builder, as well as viewing the results on the map, as this is needed to answer the question. The aim of the second task is to use the geographical search, using the map to draw a polygon around the area. The last task is aimed to force the user to use the facets, by selecting the `settlement' facet. 

        To better understand the user behaviour, we asked the participants to use the Think Aloud Protocol, as introduced in Section \ref{thinkaloud}. Specifically, we asked the participants to say what they think, see, expect, do, feel, and motivate their actions. At the end of the session we also asked the user a number of questions which can be found below.
        
        \begin{enumerate}
            \item Which elements of the system worked well?
            \item Which elements of the system did not work well?
            \item Was anything unclear?
            \item Is there any functionality that is missing, in your view?
            \item What is your opinion on the facets?
            \item What is your opinion on the map functionality?
            \item Is there anything else you would like to add to this evaluation?
        \end{enumerate}
        
        We did not include any quantitative evaluation questions\footnote{E.g. How would you rate this system on a scale from one to ten?} in the questionnaire, as satisfaction with a system was shown to be directly proportional to the quality of the results \citep{Verberne2016InformationResearch.}, and as such is not a good measure for usability. 

        To record the sessions, we used screencasts\footnote{Using the Loom Chromium plugin (\href{https://www.loom.com/}{https://www.loom.com/})}  to record the user behaviour on the screen, together with statistics on the queries recorded by the system itself. We also used sound recordings to capture the thoughts of the participants, together with notes made by the researcher (first author) sitting next to the user. 
        A table containing all queries with related statistics is available in the online data repository for this study\footnote{\url{https://doi.org/10.5281/zenodo.4064076}, also contains a list of all usability issues and a list of user needs mentioned in later sections.}.
        
        The answers to the questions, as well as the user's thoughts during searching, were transcribed and translated to English, and the resulting qualitative data were processed using grounded theory techniques \citep{Charmaz2006ConstructingAnalysis}, which entails coding statements and grouping those codes into categories, to allow for a quantitative approach on the data. 
        
        We also analysed the screencasts afterwards, and recorded all the query (re-)formulations in a pseudo-query format, together with the time spent on each query and how many results were returned.

\section{Analysis and Results}
\label{sec:results}
    
    To address our research questions, we performed both quantitative and qualitative analyses of the results of the usability evaluations. These are further detailed in the following subsections.

    A total of 148 queries were observed and recorded during the evaluation sessions, for a total of sixty-four \replaced{information needs}{research questions}, making for an average of 2.3 queries per task. A query is defined here as a combination of search terms entered into the system, an \replaced{information need}{research question} as a defined question the researcher wants to answer. The minimum number of query elements is one, as expected, and the maximum is ten, with an average of 2.4. Here, an element of a query is one AND/OR statement, so for example the pseudo query \texttt{[artefact : scraper] AND [period : neolithic]} contains two elements. 
    
    \subsection{Information Needs}   

            \begin{table*}[ht]
            \centering
            \begin{tabular}{ll}
            \textit{Find all amber from the Middle Neolithic}                                                                             &                                     \\ \hline
            \multicolumn{1}{|l|}{\textbf{Query}}                                                                                          & \multicolumn{1}{l|}{\textbf{Type}}  \\ \hline
            \multicolumn{1}{|l|}{{[}material:amber{]} AND {[}period:middle neolithic{]}}                                                  & \multicolumn{1}{l|}{}               \\ \hline
            \multicolumn{1}{|l|}{{[}material:amber{]} AND {[}period:neolithic{]}}                                                         & \multicolumn{1}{l|}{Generalisation} \\ \hline
            \multicolumn{1}{|l|}{{[}material:amber{]}}                                                                                & \multicolumn{1}{l|}{Generalisation} \\ \hline
                                                                                                                                          &                                     \\
            \textit{Find all beakers from graves in the late Neolithic}                                                                   &                                     \\ \hline
            \multicolumn{1}{|l|}{\textbf{Query}}                                                                                          & \multicolumn{1}{l|}{\textbf{Type}}  \\ \hline
            \multicolumn{1}{|l|}{{[}period:late neolithic{]} AND {[}other:grave{]} AND {[}artefact:beaker{]}}                             & \multicolumn{1}{l|}{}               \\ \hline
            \multicolumn{1}{|p{10cm}|}{{[}period:late neolithic{]} AND {[}other:grave{]} AND {[}artefact:beaker{]} AND {[}filter:prehistory{]}} & \multicolumn{1}{l|}{Specification}  \\ \hline
            \multicolumn{1}{|l|}{{[}period:late neolithic{]} AND {[}other:grave{]}}                                                       & \multicolumn{1}{l|}{Generalisation} \\ \hline
            \multicolumn{1}{|l|}{{[}period:late neolithic{]} AND {[}other:grave{]} AND {[}filter:neolitic{]}}                             & \multicolumn{1}{l|}{Specification}  \\ \hline
            
            &                                     \\
            \textit{Find all coprolites from the Swifterbant period}                                                                   &                                     \\ \hline
            \multicolumn{1}{|l|}{\textbf{Query}}                                                                                          & \multicolumn{1}{l|}{\textbf{Type}}  \\ \hline
            \multicolumn{1}{|l|}{{[}period:swifterbant{]} AND {[}artefact:coprolite{]}}                             & \multicolumn{1}{l|}{}               \\ \hline
            \multicolumn{1}{|p{10cm}|}{{[}other:swifterbant{]} AND {[}artefact:coprolite{]}} & \multicolumn{1}{l|}{Parallel / reformulation}  \\ \hline
            \end{tabular}
            \caption{Three examples of user generated tasks and their associated queries and query reformulations (translated from Dutch).}
            \label{queryexamples}
            \end{table*}
            
            Based on work on question taxonomies by \citet{Voorhees2001Overview2001} and \citet{Hermjakob2000Knowledge-basedAnswering}, we can distinguish  three main types of questions; (1) closed questions with a yes or no answer, (2) factoid questions where more than a yes/no answer is required, and (3) list questions, where a list of results is the intended end goal. Other research in the humanities such as \citet{Verberne2016InformationResearch.} suggest that humanities scholars generally have a mix of all three, with a preference for factoid questions. 
            
            In our Think Aloud sessions, we asked the users to also state the question they wanted to answer, and noted this down. We noticed that almost all the questions asked by the users are list questions, e.g. the three tasks mentioned in Table \ref{queryexamples}. 
            This intuitively makes sense for archaeologists, as research most often entails making a list of all known occurrences of a particular topic and then performing some sort of analysis on this list. In our user requirements study, the users also indicated a preference for high recall over high precision, they much prefer getting all the relevant results with some noise, than to miss some results and have only relevant results \cite{Brandsen2019}.

    \subsection{Query Strategies and Effectiveness} 
    
            We analysed the query reformulation strategies in this data, the process of altering a query to be narrower (specification, making the query longer), broader (generalisation, making the query shorter) or replacing one or more terms by other terms without making the query broader or narrower (parallel movement / reformulation)  \cite{Rieh2006AnalysisContext}. Interestingly, there is no trend to be found across all users between specification and generalisation, with both types of query reformulations occurring almost equally (twenty-five and twenty-four times, respectively). We do note that some users have a tendency to start broad and narrow down, while others do the opposite, but this seems to be a personal preference and not a preference for particular user categories. The full data is available via Zenodo\footnote{\url{https://doi.org/10.5281/zenodo.4064076}}, and there are three examples in Table \ref{queryexamples}.
            
            While the users let us know in the feedback that they liked the faceted search (see Section~\ref{evalandsatisfaction} below), when we look at the queries we see that they don't use the facets very often. Out of 148 queries, only 23 include the use of facets (15.5\%). 
            
            If we look at the use of Boolean expressions, only a small number of queries (9.5\%) use the advanced features of the query builder, i.e. have an OR or group operator. It seems that archaeologists are either not trained to think in Boolean expression, or simply do not have information needs that require them, which is in contrast with other professional search groups \citep{Russell-Rose2018}. This in turn leads to the conclusion that the query builder might be overkill for such a system, seeing as more than 90\% of the queries could have just been typed in a text field\deleted{ (with implied AND operators)}. 
            
        \subsubsection{Query Effectiveness} 

            It is difficult to directly measure the effectiveness of user queries, partly because the users themselves are not always sure that they have found the complete answer to the question. As a proxy for query effectiveness, we therefore make a comparison of the user-formulated query to a reference query of which we are sure that it returns the complete set of relevant items. The reference queries consist of query terms combined with metadata filters (facets). For example, for the task `Find all Roman pottery found in a settlement' we formulated the query \texttt{[artefact : pottery] AND [period : roman] AND [facet - site type : settlement]}. We then counted how often the users succeed in formulating the reference query. Although the users might have found the answer with a different query, this gives us an approximation of the session success.
            
            All the users managed to formulate the same query for task 1 and 2, in 1.6 and 1.2 query reformulations on average, respectively. This means they ended up using the interface in the same way as we \replaced{intended}{would}. Task 3 was more difficult, as only two out of ten participants executed a matching query. The difference in query stemmed from the confusion around the facets; we intended for the users to use the facets to filter on `settlement', but six users used `settlement' in the actual query instead and opted not to use the facets. While the facets are more exact and also handle synonyms, entering `settlement' in the query still produced \replaced{relevant}{correct} results. So even though the query was not exactly the same as the intended query, we would argue that this task was still completed by the entire user group.
            
            For the self formulated information needs, we could not determine the query effectiveness as we don't have any reference queries. Instead, we asked the users to only stop editing the query when they were satisfied with the results, and for only a couple of information needs the user indicated they were not satisfied. However, this resulted from inaccurate NER and/or documents they expected to be in the system not being present, not from the interface being difficult to use. As a quantitative approach is not possible here, we further evaluate the system using a qualitative approach in the next section.
    
    \subsection{Evaluation and User Satisfaction}
        \label{evalandsatisfaction}
        
        \subsubsection{Comments per User Group} 
        
            If we look at the number of \replaced{usability issues}{bugs and features} raised per user category (commercial, academic or government), we find that roughly 58\% of them (eighteen out of thirty-one) are raised by one user category only. 
            This indicates that it is  important to create a user group that is as diverse as possible, being representative of the target population, as otherwise certain issues will simply not be found. 
            
            \begin{figure}[b]
              \centering
                \includegraphics[width=0.5\linewidth]{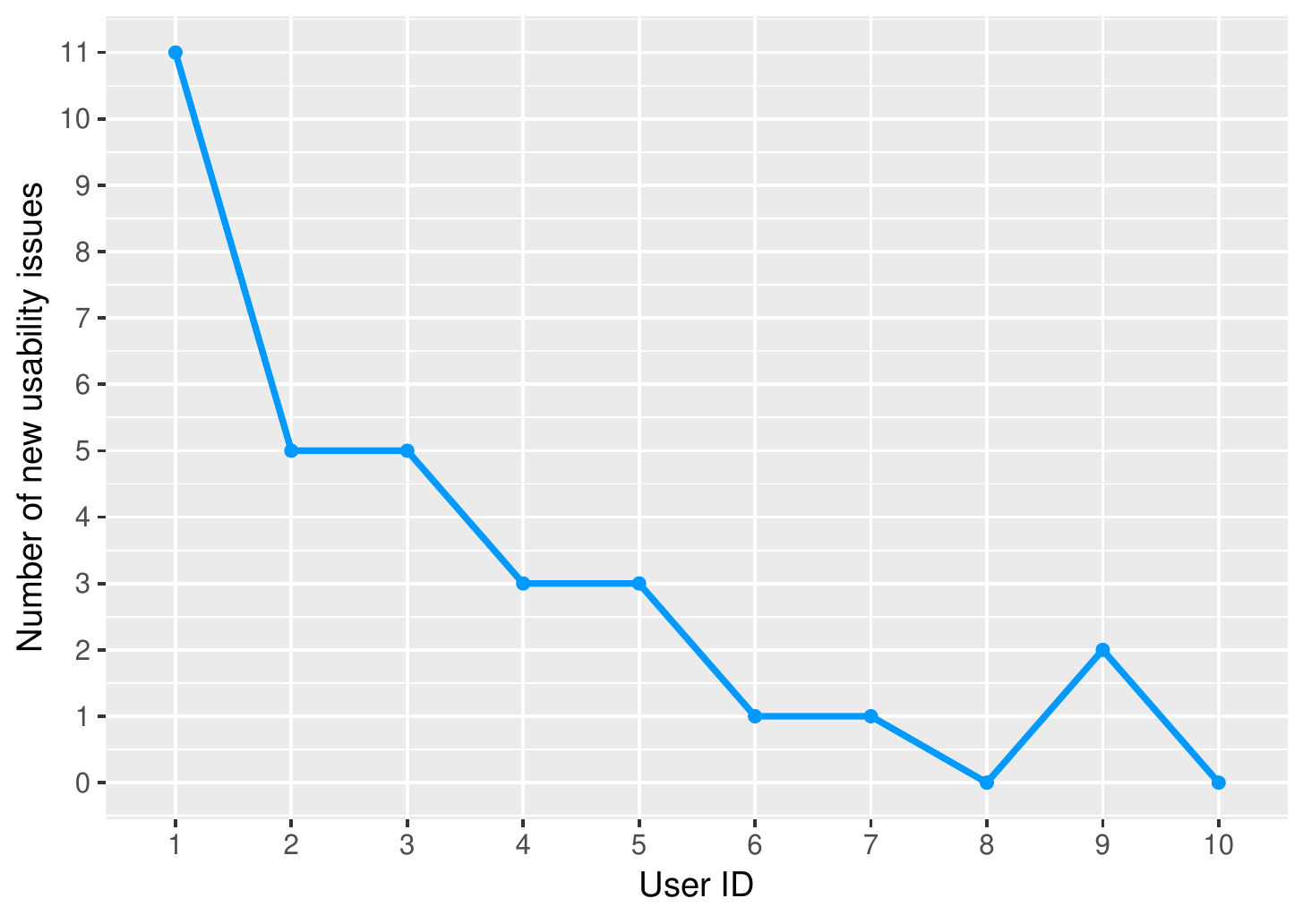}
                \caption{Line plot showing the number of new issues raised for each user} 
                \label{bugsperuser}
            \end{figure}
    
        \subsubsection{Cumulative New Issues per User}
        
            The users mentioned a total of sixty-eight usability issues, averaging 6.8 per user. Where two \added{or more} users mentioned the same issue, we grouped it, which leads to a total of thirty-one unique issues. 
            Fig. \ref{bugsperuser} shows the number of new usability issues found for each user that is added to the evaluation. We can see that after the fifth user evaluation, new users tend to not identify many new issues, confirming prior work on usability studies. The exception is user 9 with two new issues, who is the only commercial excavation archaeologist in our user group. This again underlines the necessity for a diverse user group.

        \subsubsection{Positive and Negative elements} 
        
\begin{figure}[h]
                \centering
                \includegraphics[width=0.5\linewidth]{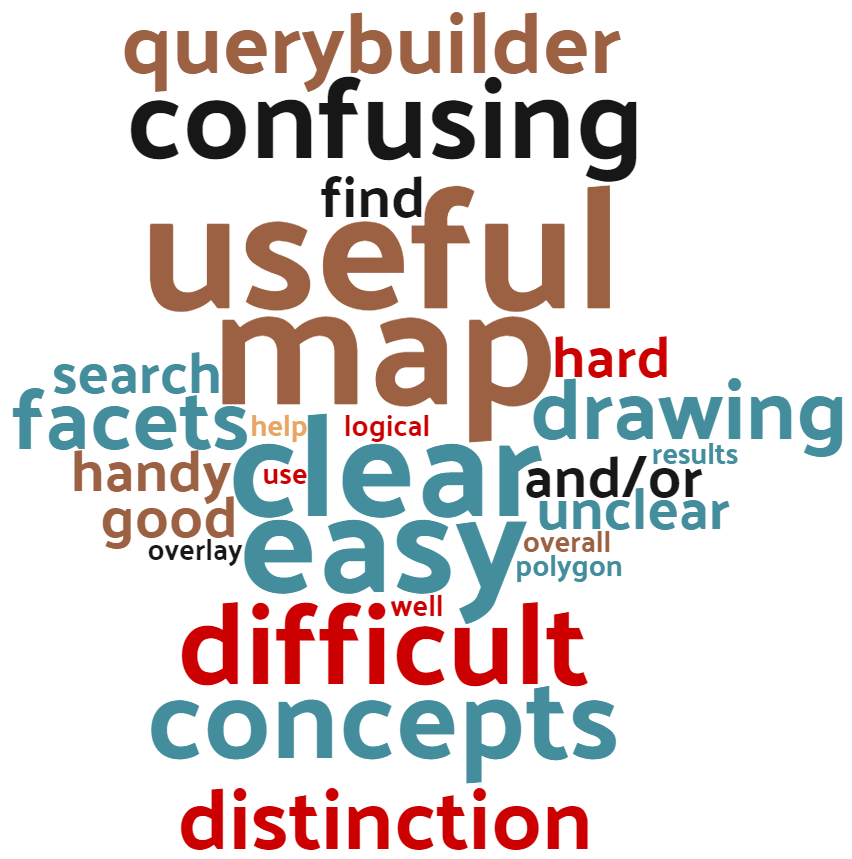}
                \caption{Word cloud of all feedback given, both positive and negative (translated from Dutch to English, `ahn' is the height model of the Netherlands)}
                \label{wordcloud}
            \end{figure}

            From the answers to the questionnaire after each session, we got the impression that overall, the users find the system fairly easy to use and clear. The map functionality is mentioned by everyone, and mentioned often, something that was expected by the results of the user requirement solicitation. In Fig. \ref{wordcloud} we have plotted a word cloud of all feedback, after translation from Dutch to English. We lowercased all text, removed punctuation, removed stopwords (NLTK list), and then plotted only words which occurred more than once.
            
            We can see that the words `clear' and `easy' are often used, as well as the map, confirming the impression we got from the sessions. Also we see the words `difficult' and `unclear' used often, these are more in relation to negative aspects of the system. 
            
            In table \ref{posnegfeedback} we show the most frequent words for the positive and negative feedback fields, respectively, where we have removed all stop words, verbs and opinion-bearing words (such as `clear', `hard'). Again we only include words mentioned more than once.

            \begin{table}[t]
                \centering
                \begin{tabular}{|l|l|l|l|}
                \hline
                \multicolumn{2}{|c|}{\textit{Positive}} & \multicolumn{2}{c|}{\textit{Negative}} \\ \hline
                \textbf{Freq.}      & \textbf{Word}     & \textbf{Freq.}     & \textbf{Word}     \\ \hline
                9                   & map               & 6                  & concepts          \\ \hline
                4                   & querybuilder      & 5                  & facets            \\ \hline
                3                   & facets            & 4                  & and/or            \\ \hline
                3                   & usability         & 3                  & intuitiveness     \\ \hline
                2                   & drawing           & 2                  & help              \\ \hline
                2                   & overall           &                    &                   \\ \hline
                \end{tabular}
                \caption{Feedback split into positive and negative, with for each word how often it occurs in that context. Words only mentioned once are not included.}
                \label{posnegfeedback}
            \end{table}

            On the negative side we can see that choosing which concept to search for, intuitiveness, the `help' button, the facets, and the AND/OR toggle buttons are elements that are commonly experienced as negative at the moment. These issues and features will be dealt with in the next version of the system. We also see that the map, query builder, facets, and overall usability are often experienced as positive.
            
            One of the other observations made during the evaluation is that none of the users use, or even see, the `Help' button, which we did expect them to. This led to some preventable confusion about the system, as some questions the users had were actually explained in the help section. As a solution, we will include in-context help in the next version; pop ups that appear when hovering on certain elements to further explain the system.
   
        \subsubsection{Time Spent per Query}
        
\begin{figure}[t]
              \centering
                \includegraphics[width=0.5\linewidth]{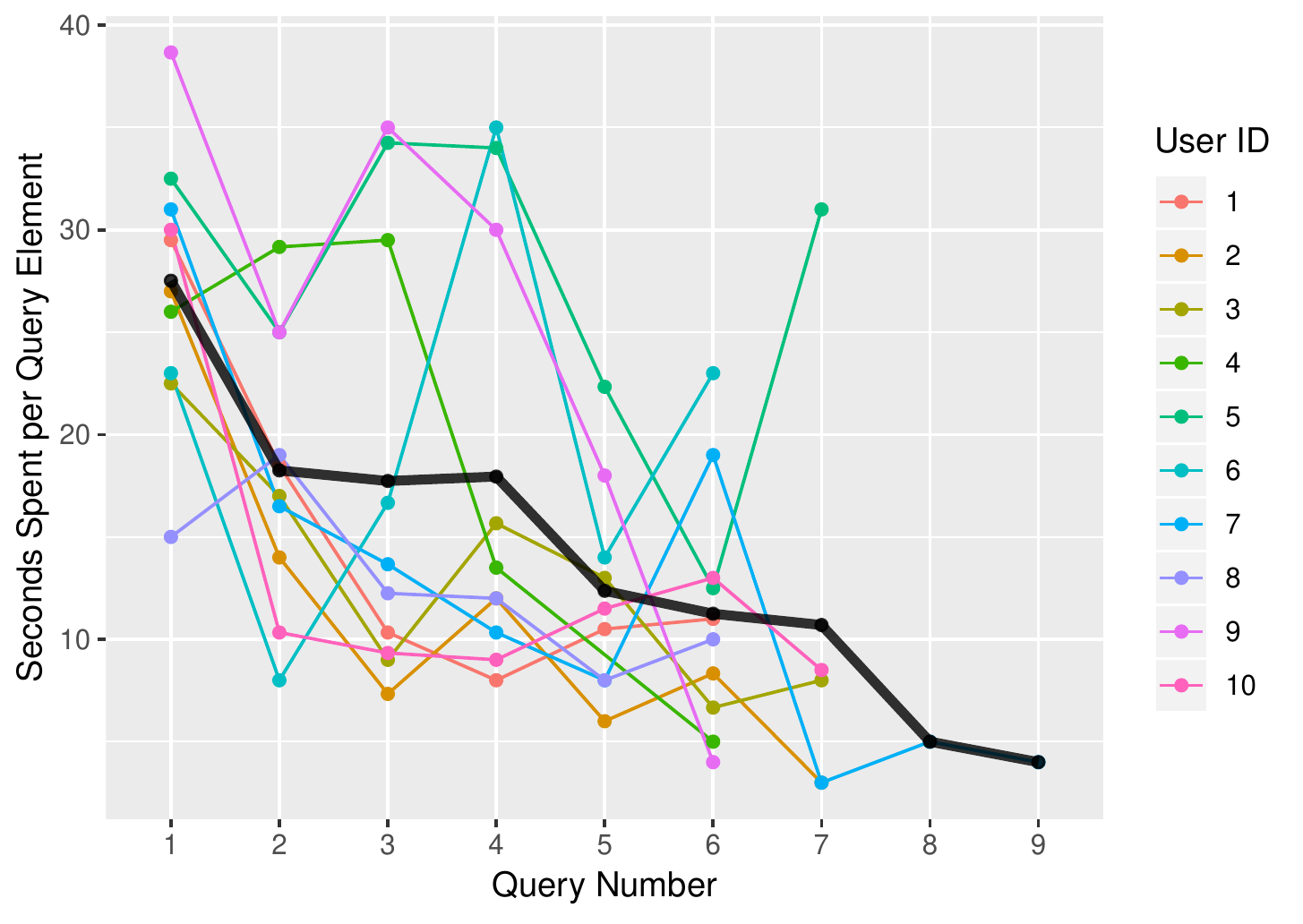}
                \caption{Line plot showing for each user, how much time they spent formulating one element of a query, for each new query they attempted. The black line is the average over all the users.}
                \label{lineplot}
            \end{figure}

            As mentioned before, we observed sixty-four research questions, with a total of 148 queries, so 2.3 query reformulations on average. For each initial query and query reformulation, we recorded the time taken to (re-)formulate the query and the number of elements in the query, among other information. We use the time per element instead of time per query to account for the difference in length of query between users, this way we can easily compare them. In Fig. \ref{lineplot} we plotted the time per element against the succession of queries attempted by a user. Here we see a clear downward trend (average between users shown in black). As the users had to do at least three of their own \replaced{tasks}{queries}, but could continue with more if they wanted, means that we have less data between query 6 and 9. However the trend is already clear between query 1 and 6.
            
            This trend means that as the users perform more queries, the time taken per query element decreases rapidly, indicating that the system is easy to learn. 

\section{Discussion}
\label{sec:discussion}

    \citet{Gibbs2012BuildingDesigns} talk about how the typical humanities user is often neglected in the design of tools, and how tools' visibility can be increased by good attention to the usability. More recent work by \citet{Bulatovic2016UsabilityProcessb} agrees, and states that digital humanities tools often suffer from poor user experiences, mainly caused by the lack of resources spent on usability research.  
    
    As mentioned in the introduction, it is important to evaluate usability, as previous research indicates that usability and uptake of search systems is strongly correlated \citep{Dudek2007IsEngines}. At the same time, it is difficult to evaluate usability independently from the quality of the results, as users will perceive a system as not being usable if the results they get are of low quality. In this work, we found that it is important to brief the test users before hand to manage their expectations, and design the tasks and questionnaire to specifically target usability features that can be evaluated whether the results are good or bad.
    
    \citet{Bulatovic2016UsabilityProcessb} also mention that early iterative cycles of testing should be implemented in these kinds of projects, to avoid common usability problems. This is what we are doing in this project, and we hope this will see more uptake in the digital humanities, as it seems usability evaluation is something done at the end of most projects as an afterthought, if done at all. In 2012, \citeauthor{Gibbs2012BuildingDesigns} called for a shift to user-centred design techniques, and luckily we do see that most of the more recent studies take this approach \citep[e.g.][]{Hinrichs2015TradingExploration,vanZundert2016TheEdition,Esmailpour2019DevelopingQueries}.
    
    We think that for tools to be used by humanities scholars, the user interface needs significant investment in the design that needs to be integrated into the project budget and timeline. At a more broader level, we agree with \citet[][p. 20]{Koolen2018TowardPractice} that digital tools `always require critical reflection on how they mediate between researchers and their materials of study', something we will investigate further in future research.
    
    Specifically for the archaeology domain, usability is evaluated and published even less than in the digital humanities as a whole. Seeing as there are key limitations to the currently available systems, and usability studies are rare in the archaeology domain, we think it is vital to research and publish the usability of the system we are creating. More generally, we believe that the research presented here is not only of value to the system itself, but also to other researchers building online tools; perhaps the findings are not generalisable to other applications due to the small sample size, but can at the very least serve as inspiration. When more archaeologists publish their usability studies, we can together make more useful, meaningful tools.
        
\section{Conclusions} 
\label{sec:conclusions}
    
    In this paper, we have investigated how Dutch archaeologists prefer to use online search, what features they deem positive and negative, how well our UI performs, and have assessed which analyses are useful for usability studies of this and similar systems. Here we will answer our research questions.
    
    \paragraph{1. What type of information needs do archaeologists have?}
    
        From previous studies and our own user requirement solicitation study, we see that Dutch archaeologists are mainly interested in geographic search, plotting results on a map, and faceted search.
         
        We see that Dutch archaeologists have a clear preference for high-recall list-type questions when doing research. 
        \replaced{A}{Another} difference between archaeologists and other professional search domains \citep{Russell-Rose2018} is the lack of preference for Boolean expressions, our user group barely used them, nor told us they wanted them.
    
    \paragraph{2. What query strategies do archaeologists use?}
    
        We did not find any preference on query reformulation: specification and generalisation occurred roughly equally. 
        We also noted that all users were able to create the reference queries for the predefined tasks, indicating the UI is being used as we intended.
        Regarding facets, we see that while users report these as being helpful, they do not use them very often, occurring in only 15\% of the queries.
        
    \paragraph{3. How satisfied are the users with the usability of the AGNES system?}
    
        By analysing the feedback during the system evaluation, we found that users found the UI easy to use, clear, and useful. They specifically found the map features and query builder to be good features of the system. When we visualised the feedback, we see that the query builder, map features, facets, and snippets are experienced as positive. Some negative features include the help button, uncertainty about the mechanism behind the facets and concepts in the query builder, and the overall intuitiveness. 
    
        We see that the time taken per query element decreases fairly rapidly when users perform more queries, which indicates the system is easy to learn.
        
        Using a relatively small user group of ten participants was expected to be enough to find and address usability issues, and this proved to be correct; we found that as the number of users increased beyond five, the number of issues highlighted dropped rapidly. 
        
        The importance of a diverse user group has been shown, as we found that roughly two thirds of issues were only raised by one of the user groups. Interestingly, if this is combined with the previous conclusion, this might mean that the ideal size of a user group might be five users per user category, instead of five in total.

        \bigbreak   
            
        In conclusion, it seems that AGNES can address the problem of accessing grey literature in Dutch archaeology, although this needs to be evaluated more thoroughly by comparing the results found with the use of AGNES to the prior knowledge of the topic, i.e. lists of occurrences of certain types of artefacts archaeologists have compiled manually. We are hopeful that AGNES will help archaeologists to answer their research questions more effectively and efficiently, leading to a more coherent narrative of the past.

\subsection{Future Work}
	\label{futurework}

	The work discussed in this paper is the result of the second year of a four year project. Each year, a new version of AGNES is developed, tested, and evaluated by the focus group. The first two workshops dealt with user requirement solicitation and evaluation of the interface, for the next workshop we will evaluate the quality of the results returned.
	
	Further work is needed to refine the user interface, all the issues and suggestions raised by the user group will be dealt with in the next version of the system. This should make it easier to focus purely on evaluating the results in the next workshop.
	
	At the moment, we only evaluated the system using ten users. We believe that a quantitative study using statistics generated by the system could be useful in finding usability issues, as well as seeing patterns in usage. To this end we will make the next version of the system public and invite a large group of archaeologists to use the system. This should give us a much larger user group, although this is a more superficial analysis and loses some of the depth of evaluations done on the current group with the one-on-one approach. 
	
	Some recent work by \cite{Russell-Rose2020DesigningParadigm} suggests that traditional query builders like the one used in this project might not be ideal, and a more visual layout of a query provides a more direct mapping to the underlying semantics, and makes it more transparent. This is something we'd like to experiment with in future versions of AGNES, especially since our user group didn't seem to need the query builder for boolean expressions.

\clearpage 


\clearpage 

\doublespacing        
\bibliographystyle{dsh}
\bibliography{offline-copy}

\end{document}